\documentclass[aps,prl,english,twocolumn,superscriptaddress]{revtex4-1}
\usepackage[T1]{fontenc}
\usepackage[latin9]{inputenc}
\usepackage{units}
\usepackage{graphicx,color}
\usepackage[normalem]{ulem}
\usepackage{hyperref}
\usepackage{bm}

\makeatletter
\usepackage{breqn}

\gdef\wrap@breqn@environ#1#2{
    \expandafter\let\csname breqn@oldbegin@#1\expandafter\endcsname\csname #1\endcsname
    \expandafter\let\csname breqn@oldend@#1\expandafter\endcsname\csname end#1\endcsname
    \expandafter\gdef\csname breqn@begin@#1\endcsname{%
        \expandafter\let\csname #1\expandafter\endcsname\csname breqn@oldbegin@#1\endcsname%
        \begin{#2}%  
    }
    \expandafter\gdef\csname breqn@end@#1\endcsname{%
        \expandafter\let\csname end#1\expandafter\endcsname\csname breqn@oldend@#1\endcsname%
        \end{#2}%
        \expandafter\let\csname #1\expandafter\endcsname\csname breqn@begin@#1\endcsname%
        \expandafter\let\csname end#1\expandafter\endcsname\csname breqn@end@#1\endcsname%
    }
    \expandafter\let\csname #1\expandafter\endcsname\csname breqn@begin@#1\endcsname
    \expandafter\let\csname end#1\expandafter\endcsname\csname breqn@end@#1\endcsname
}

\wrap@breqn@environ{equation}{dmath}
\wrap@breqn@environ{equation*}{dmath*}

\makeatletter
\let\cat@comma@active\@empty
\makeatother

\makeatother

\usepackage{babel}
\begin{document}

\title{Direct Measurement of Anharmonic Decay Channels of a Coherent Phonon}

\author{Samuel W. Teitelbaum}

\affiliation{PULSE Institute of Ultrafast Energy Science, SLAC National Accelerator
Laboratory, Menlo Park, California 94025, USA}

\affiliation{Stanford Institute for Materials and Energy Sciences, SLAC National
Accelerator Laboratory, Menlo Park, California 94025, USA}
\email{steitelb@slac.stanford.edu}

\selectlanguage{english}

\author{Tom Henighan}

\affiliation{PULSE Institute of Ultrafast Energy Science, SLAC National Accelerator
Laboratory, Menlo Park, California 94025, USA}

\affiliation{Department of Physics, Stanford University, Stanford, California
94305, USA}

\author{Yijing Huang}

\affiliation{PULSE Institute of Ultrafast Energy Science, SLAC National Accelerator
Laboratory, Menlo Park, California 94025, USA}

\affiliation{Department of Applied Physics, Stanford University, Stanford, California
94305, USA}

\author{Hanzhe Liu}

\affiliation{PULSE Institute of Ultrafast Energy Science, SLAC National Accelerator
Laboratory, Menlo Park, California 94025, USA}

\affiliation{Department of Physics, Stanford University, Stanford, California
94305, USA}

\author{Mason P. Jiang}

\affiliation{Department of Physics, Stanford University, Stanford, California
94305, USA}

\affiliation{PULSE Institute of Ultrafast Energy Science, SLAC National Accelerator
Laboratory, Menlo Park, California 94025, USA}

\author{Diling Zhu}

\affiliation{LCLS, SLAC National Accelerator Laboratory, Menlo Park, California
94025, USA}

\author{Matthieu Chollet}

\affiliation{LCLS, SLAC National Accelerator Laboratory, Menlo Park, California
94025, USA}

\author{Takahiro Sato}

\affiliation{LCLS, SLAC National Accelerator Laboratory, Menlo Park, California
94025, USA}

\author{\'{E}amonn D. Murray}

\affiliation{Department of Physics and Department of Materials, Imperial College
London, London SW7 2AZ, United Kingdom}

\author{Stephen Fahy}

\affiliation{Tyndall National Institute, Cork, Ireland}

\affiliation{Department of Physics, University College Cork, Cork, Ireland}

\author{Shane O'Mahony}

\affiliation{Tyndall National Institute, Cork, Ireland}

\affiliation{Department of Physics, University College Cork, Cork, Ireland}

\author{Trevor P. Bailey}

\affiliation{Department of Physics, University of Michigan, Ann Arbor, Michigan
48109, USA}

\author{Ctirad Uher}

\affiliation{Department of Physics, University of Michigan, Ann Arbor, Michigan
48109, USA}

\author{Mariano Trigo}

\affiliation{PULSE Institute of Ultrafast Energy Science, SLAC National Accelerator
Laboratory, Menlo Park, California 94025, USA}

\affiliation{Stanford Institute for Materials and Energy Sciences, SLAC National
Accelerator Laboratory, Menlo Park, California 94025, USA}

\author{David A. Reis}

\affiliation{PULSE Institute of Ultrafast Energy Science, SLAC National Accelerator
Laboratory, Menlo Park, California 94025, USA}

\affiliation{Stanford Institute for Materials and Energy Sciences, SLAC National
Accelerator Laboratory, Menlo Park, California 94025, USA}

\affiliation{Department of Applied Physics, Stanford University, Stanford, California
94305, USA}

\affiliation{Department of Photon Science, Stanford University, Stanford, California
94305, USA}
\begin{abstract}
We observe anharmonic decay of the photoexcited coherent $A_{1g}$ phonon in bismuth to points in the Brillouin zone where conservation of momentum and energy are satisfied for three-phonon scattering.   The decay of a coherent phonon can be understood as a parametric resonance  process whereby the atomic displacement periodically modulates the frequency of a broad continuum of modes. This results in energy transfer through resonant squeezing of the target modes. Using ultrafast diffuse x-ray scattering, we observe build up of coherent oscillations in the target modes driven by this parametric resonance over a wide range of the Brillouin zone.  We compare the extracted anharmonic coupling constant to first principles calculations for a representative decay channel.

\end{abstract}

\date{\today}

\maketitle 

Lattice anharmonicity governs a broad range of phenomena in condensed matter physics, from structural phase transitions \cite{Zhong1995} to heat transport and thermoelectricity \citep{Kittel:ISSP,Zebarjadi2012}. 
Recent advances in first-principles calculations have allowed for precise calculations of  thermal properties including effects due to phonon-phonon scattering~\citep{Debernardi1995,Broido2007,Togo2015}. However, experimental validation is indirect, mostly limited to bulk properties like the thermal conductivity, and experimental probes of the microscopic details of anharmonicity remain elusive. While scattering techniques like inelastic neutron scattering \citep{Brockhouse1955,lovesey1971theory,squires1978introduction} and inelastic x-ray scattering \citep{Krisch2007,burkel2000phonon,Baron2014,Baron2014a} can measure quasi-harmonic properties of phonons such as frequencies and linewidths across the Brillouin zone, these techniques lack the ability to resolve the individual decay channels of a given mode. Such a measurement would provide a unique view of the processes occurring during thermal equilibration, including a quantitative measurement of anharmonic force constants.  

Here, we report direct measurements of the anharmonic coupling of
the zone center Raman active $A_{1g}$ optic phonon in bismuth to longitudinal acoustic phonons at high wavevector, and extract an anharmonic coupling constant for a subset of these modes that is within an order of magnitude of that obtained by first-principles calculations.  These measurements utilize femtosecond  x-ray diffuse scattering to probe the temporal evolution of the phonon mean square displacements following optical excitation \citep{Trigo2013}.  It was proposed in \citep{Fahy2016} that a parametric resonance of the zone-center mode with the acoustic branch would be observable in femtosecond scattering measurements using an x-ray free electron laser (FEL).  

\begin{figure}[h]
\includegraphics{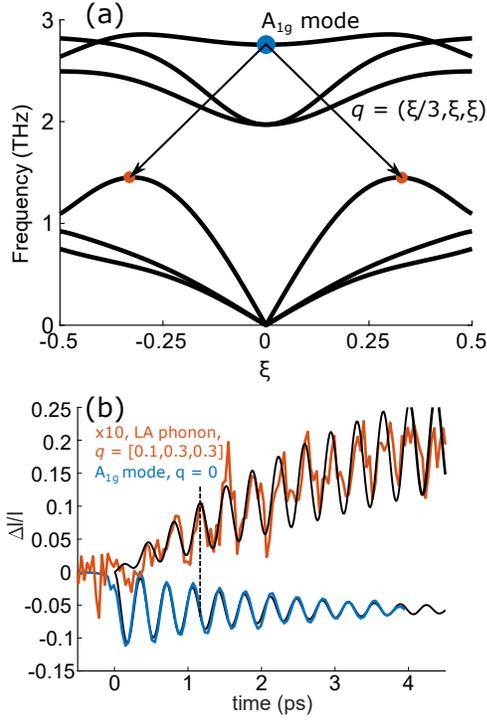}
\caption{Color online. (a) Phonon dispersion relation in bismuth along the
$\bm{q}=(\frac{1}{3}\xi\ \xi\ \xi)$ direction, illustrating a decay
channel of an $A_{1g}$ phonon into a pair of LA phonons at $\bm{q}$ and
$-\bm{q}$. (b) experimental signature of decay of the $A_{1g}$ phonon
in bismuth by the channel shown in (a). The lower (blue) curve shows
the relative intensity change of the (2 3 2) Bragg peak, which is proportional to the $A_{1g}$ mode amplitude. The upper (orange)
curve shows the relative intenisty change of the diffuse scattering
$\Delta I/I$ in a region near $\bm{q}=(0.1\ 0.3\ 0.3)$ in the (0 1 1)
zone (multiplied by 10). The black lines are simulations. The dashed
line indicates a $\pi/2$ phase shift between the $A_{1g}$ mode and the
target mode. \label{fig:Major-Results}}
\end{figure}
 
Bismuth is a group V semimetal that exhibits particularly strong electron-phonon
coupling due to its Peierls distored structure, making it an ideal testbed for the study of large-amplitude
phonon motion. 
Upon photoexcitation, the sudden change in the lattice potential causes the atoms to move
coherently along the $A_{1g}$ mode coordinate, resulting in a macroscopic
modulation of both the optical reflectivity \citep{Cheng1990} and the Bragg peaks that
are sensitive to the $A_{1g}$ structure factor modulation \citep{Sokolowski-Tinten2003,Fritz2007,Johnson2009}.

We use the theoretical framework developed in reference \citep{Fahy2016} to
describe the  anharmonic coupling of the photoexcited coherent phonon to pairs of phonons with
large wavevector (target modes), in a process analogous to optical parametric downconversion (as shown schematically in Fig.~\ref{fig:Major-Results}(a)). 
The Hamiltonian describing cubic coupling of a single
mode with normal mode amplitude and momentum $(U_{0},P_{0})$ to a branch of $N$ modes with reduced wavevector $\bm{q}$ can be expressed as
\begin{equation}
H=\frac{1}{2}\left[P_{0}^{2}+\Omega^{2}U_{0}^{2}\right]+\frac{1}{2N}\sum_{\bm{q}}\left[p_{\bm{q}}^{2}+\omega_{\bm{q}}^{2}(1+2g_{\bm{q}}U_{0})u_{\bm{q}}^{2}\right]\label{eq:Hamiltonian}
\end{equation}
where $p_{\bm{q}}$ and $u_{\bm{q}}$ are the corresponding normal mode
momentum and displacement and $g_{\bm{q}}$ specifies the strength of the anharmonic coupling  between the zone-center phonon and phonons at $\pm \bm{q}$.  
Note that a coherent zone-center phonon has $\langle U_{0} \rangle \sim \cos(\Omega t)$ such that it parametrically modulates the frequency of the other phonons, driving squeezing oscillations in their mean square displacements \citep{Xin1989,Fahy2016}
\begin{equation}
\Delta\left\langle u_{q}^{2}(t)\right\rangle =\frac{\mp kTg_{q}A}{
\omega_{q}\sqrt{(\gamma^\prime)^{2}+(2\omega_{q}-\Omega)^{2}}}
\left[e^{-\gamma_{0}t/2}\sin(\Omega t+\delta')-e^{-\gamma_{q}t}\sin(2\omega_{q}t+\delta')\right],\label{eq:Squeezing}
\end{equation}
where $A$  and $\gamma_{0}$ are the amplitude and (energy) damping rate of the coherently excited driving mode at frequency $\Omega$ $\omega_{\bm{q}}$ and $\gamma_{\bm{q}}$ are the frequency and damping rate of the target mode at $q$.  $\gamma^\prime = \gamma_q-\gamma_0/2$ and $\delta'=\tan^{-1}[(2\omega_{\bm{q}}-\Omega)/(\gamma_{\bm{q}}-\gamma_{0}/2)]$. The negative (positive) sign corresponds to  $\gamma^\prime > 0 \ (\gamma^\prime < 0)$, and the ``$\Delta$'' means that we have subtracted off the thermal equilibrium value before excitation.  
Note the squeezing is most effective when the  parametric resonance condition is met, \emph{i.e.}  $\Omega=2\omega_{\bm{q}}$.  More generally, non-degenerate coupling to different branches ($i$ and $j$) is also possible. In this case the parametric resonance condition is $\Omega = \omega_{\bm{q}i}\pm\omega_{\bm{q}j}$.  In Eq.~\ref{eq:Squeezing}, we have taken the target mode occupation to be classical, which is appropriate for
bismuth at room temperature.

The experiment was carried out using the XPP instrument at the LCLS x-ray FEL
with a photon energy of 9.5 keV selected using a diamond double-crystal
monochrometer \citep{Chollet2015}. The sample was rotated such that the x rays propagated at a 71 deg. angle with respect to $(2 \bar{1} \bar{1})$ (binary axis). 800 nm, $\sim 45 \  \mathrm{fs}$ pump pulses were focused onto a 50 nm thick (111) epitaxial Bi film on $\mathrm{BaF_2}$ at a 1.8 degree angle of incidence with respect to the surface with an incident fluence of $2.5 \ \mathrm{mJ/cm^2}$. The x-ray pulses were less than 50 fs in duration and contained $\sim 10^9$ photons per shot  at a repetition rate of 120 Hz. The x rays were incident on the sample at an angle of 0.5~degrees relative to the surface so that their penetration depth matched the thickness of the Bi film.  The delay between the optical and x-ray pulses was controlled using a fast-scan delay stage, and the fine timing was measured on a single-shot basis using the XPP timing tool~\citep{Chollet2015}. The overall time resolution of the instrument is better than 100~fs, allowing observation of oscillations in the x-ray intensity to $\sim$5~THz, more than sufficient to measure the coherent $A_{1g}$ mode at 2.85 THz. A polycrystalline $\mathrm{LaB_{6}}$ sample was used for calibration of the pixel array detector  (CSPAD \citep{Herrmann2013}) position used to simultaneously collect scattered x rays over a wide range of momentum transfer, $\bm{Q}$. 

\begin{figure*}[t]
\includegraphics{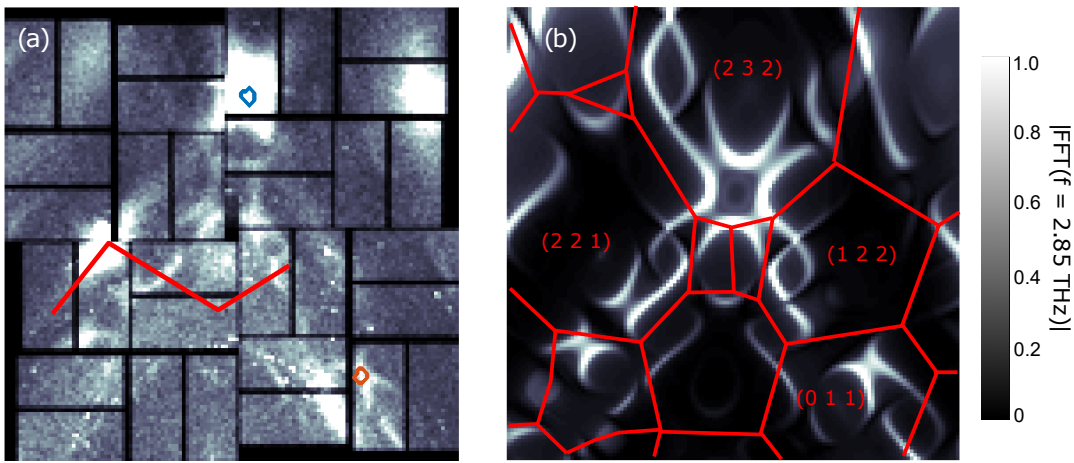}
\caption{Color online. (a) Intensity map of the Fourier transform of the femtosecond diffuse scattering signal at $f=2.85\mathrm{\ THz}$.
The red line is the path of the lineout shown in Fig.~\ref{fig:Lineouts}.
The blue and orange boxes outline the regions used to extract the
time-domain traces in Fig.~\ref{fig:Major-Results}. The four brightest
spots in (a) are due to oscillations of the structure factor from
the $A_{1g}$ mode near Bragg peaks and are therefore not present
in (b). Brillouin zones relevant to the text are labeled in (b). (b)
Predicted intensity map using the dispersion relation of bismuth and
the parametric resonance condition, with a uniform anharmonic coupling
constant $g=1.0$. The red lines in (b) show the Brillouin zone boundaries.\label{fig:3Thz-image}}
\end{figure*}

A major result of our study is shown in Fig.~\ref{fig:Major-Results}. Fig.~ \ref{fig:Major-Results}(a) shows the calculated phonon dispersion of bismuth along the $\bm{q}=(\xi/3\ \xi\ \xi)$ direction.
In Fig.~\ref{fig:Major-Results}(b) the blue curve shows the relative
intensity change $\Delta I/I$ near the $(2\ 3\ 2)$ Bragg peak, which oscillates in time with the coherent $A_{1g}$ mode displacement. The intensity oscillation near the Bragg peak is fit to a decaying cosine function, which is used to extract the oscillation amplitude of the $A_{1g}$ mode from the structure factor. The upper plot (orange curve) represents the relative intensity change (multiplied by 10) around
$\bm{q}=(0.1\ 0.3\ 0.3)\ $ 
reciprocal lattice units (r.l.u.), where the
(highest) acoustic phonon frequency $\omega_{\bm{q}}$ is close to half of $\Omega_{A1g}$.
The intensity oscillations in this region  are attributed to resonantly
squeezed phonons with predominantly LA character driven by anharmonic
coupling with the $A_{1g}$ mode. The black curves show simulation
results for the diffuse scattering in a region around $\bm{Q} = (0.1 \ 1.3 \ 1.3)$ r.l.u., derived
from Eq.~(\ref{eq:Squeezing}), which also includes a slow increase of the
diffuse scattering to account for the slow heating of the lattice,
and a fit to the Bragg peak time dependence due to modulation of the
$A_{1g}$ mode displacement. From the $A_{1g}$ mode oscillation, we extract a softened
frequency of $2.82\ \mathrm{THz}$, and an amplitude of $A=9.74\times10^{-4}$
(in units of the c-axis of bismuth), or $1.15\ \mathrm{pm}$, and
a decay rate of $\gamma_{0}=0.82\ \mathrm{ps^{-1}}$. The extracted amplitude is in good agreement with previous measurements with similar bond softening \cite{Fritz2007}.

The only additional adjustable parameters in our model for the amplitude
of the oscillations $\left\langle u_{\bm{q}^{2}}\right\rangle$ are the anharmonic coupling constant $g_{\bm{q}}$
and target mode lifetime $\gamma_{\bm{q}}$. Therefore, this experiment
allows us to obtain the $A_{1g}-LA$ phonon coupling constant by measuring
the amplitude of the Bragg peak modulation due to the $A_{1g}$ mode,
and $\left\langle u_{\bm{q}^{2}}\right\rangle$, given the phonon lifetime $\gamma_{q}$.
Assuming $\gamma_{0}\gg\gamma_{\bm{q}}$, perfect resonance, and that the observed mode dominates the static diffuse scattering, we can put a lower bound on the coupling strength with no additional input from theory. Under these assumptions, we extract
$g_{\bm{q}}=-0.7$ for the LA mode shown in Fig.~\ref{fig:Major-Results}.  An amplitude of 1.0 means that for a 1\% displacement in the position of the bismuth atom along the A1g mode (relative to the c-axis) there is a 1\% change in the LA phonon frequency. Here we have chosen a convention whereby the negative sign means that as the atom moves toward the center of the unit cell ($U_0$ negative), the LA phonon mode hardens.

We can make a more sophisticated estimate using theoretical predictions for the phonon dispersion and eigenvectors based on harmonic forces.  For a particular mode on branch $i$, the partial contribution to the scattered intensity at scattering vector $\bm{Q}$ is
\begin{equation}
I_i(\bm{Q}) \propto \left\langle u_{\bm{q},i}^{2}\right\rangle  \left| \sum_{s}{\bm{Q}\cdot\bm{\epsilon}^{(s)}_{\bm{q},i} e^{i\bm{Q}\cdot \bm{r}_s}}\right|^{2}
\label{eq:DS-intensity}
\end{equation}
Here  $\bm{r}_s$ is the equilibrium position and $\bm{\epsilon}^{(s)}_{\bm{q},i}$ is the eigenvector corresponding to the $s$th atom in the unit cell.

The  measured value of the coupling constant depends on the amplitude and decay rate of the zone center mode, the decay rate of the target mode and the contribution to the scattered intensity from the target mode, $ I_i(\bm{Q})\propto\langle u_{\bm{q},i}^2 \rangle$. 
The uncertainty in the measured value of the coupling constant is primarily due to systematic errors.  These errors are associated with our estimate of the decay rate of the target mode $\gamma_{\bm{q}}$, and our extraction of the scattered intensity from the squeezed mode, $ I_i(\bm{Q})$, from $\Delta I/I$.  As described below, we estimate $\gamma_{\bm{q},i} =~\mathrm{ 0.3~ps}^{-1}$ based on the damping rate for acoustic modes near 1 THz (generated by sudden squeezing due to  photoexcitation).  The measured value of $\langle u_{\bm{q},i}^2 \rangle$ depends on the ability to separate the contribution from a single mode from other sources of  scattering at $\bm{Q}$, including diffuse scattering from other phonons as well as other sources (\emph{e.g.} due to static disorder and Compton scattering).  

Assuming diffuse scattering from phonons dominates $I(\bm{Q})$, we can separate  contributions from the individual branches using Eq.~(\ref{eq:DS-intensity}).  Density functional perturbation theory (DFPT) calculations were used
to determine the mode frequencies and eigenvectors in the region shown as black curves
in Fig.~\ref{fig:Major-Results}(a). Using these frequencies and eigenvectors,  our estimate for the LA phonon decay rate,
and integrating over the signal in the region on the detector outlined in orange in Fig.~\ref{fig:3Thz-image}(a), we extract an anharmonic coupling constant of $g_{\bm{q}}= -1.0$ for $\bm{q}=(0.1\ 0.3\ 0.3)$ r.l.u. DFT frozen-phonon calculations predict a value of $g_{\bm{q}} = -6$ in this region of reciprocal space, which is within an order of magnitude of our experimentally measured value. When summing over the Brillouin zone, the calculations predict an $A_{1g}$  phonon decay rate of 0.34 ps$^{-1}$ \cite{Fahy2016} at room temperature, in reasonable agreement with the experimentally determined decay rate at low fluence of 0.5 ps$^{-1}$ \cite{Li2013}, and 0.82 ps$^{-1}$ for the excitation conditions used in this work.

For a fixed experimental geometry, we are able to observe the anharmonic decay to modes covering a large portion of the Brillouin zone.  
In order to identify these modes, we  look for regions of diffuse scattering that show intensity oscillations at $f_{A1g}=2.85\mathrm{\ THz}$.
This is accomplished by taking a Fourier transform along the time
axis for each detector pixel, and plotting the intensity of the Fourier
transform at the frequency frame closest to the $A_{1g}$ mode frequency. The resulting intensity map is shown in Fig.~\ref{fig:3Thz-image}(a).
The delay was scanned up to 4.8 ps after the arrival of the optical
pulse, so the frequency map has a bandwidth of 0.2 THz, approximately
one quarter the linewidth of the $A_{1g}$ mode. 
Figure~\ref{fig:3Thz-image}(b) shows the calculation of the Fourier intensity at 2.85 THz, assuming the mode contribution to the intensity is given by Eq.~\ref{eq:DS-intensity} with $\langle u^2_{\bm{q},i}(t)\rangle$ given by Eq.~\ref{eq:Squeezing} and with the eigenvectors and frequencies computed from DFPT. This calculation reproduces well the regions of Q-space where the resonance condition $2\omega_{\bm{q},i} = \Omega_{\rm A1g}$ is achieved.

Impulsive optical excitation of hot carriers
drives the coherent $A_{1g}$ phonon \citep{Fritz2007,Cheng1990}
as well as a continuum of squeezed modes across the Brillouin zone by a second-order Raman-like process~\cite{Trigo2013,Henighan2016,Johnson2009}. 
These two effects manifest themselves differently
in our data. While the coherent $A_{1g}$ phonon modulates the structure
factor for wavevectors near the zone center at $\Omega$, the squeezed
modes modulate $\left\langle u_{\bm{q}}^{2}\right\rangle $ and thus the
intensity oscillates at $2\omega_{\bm{q}}$ across the Brillouin zone,
mostly away from $q=0$ \citep{Trigo2013,Zhu2015,Henighan2016}.
Hereafter, we refer to this effect as \emph{sudden squeezing}.
In addition, the coherent $A_{1g}$ mode resonantly drives the mean-squared
displacements of modes at $2\omega_{\bm{q}}\approx\Omega$, which results
in an additional oscillatory component at $2\omega_{\bm{q}}$, 
referred hereafter as \emph{resonant squeezing}.

Several features allow us to distinguish the effects of sudden squeezing from resonant squeezing on $\left\langle u_{\bm{q}}(t)^{2}\right\rangle$. 
First, the differential change in the mean-square phonon displacements due to the oscillation of the coherent $A_{1g}$ mode should start near zero at $t=0$
and build up slowly as energy is transferred from the $A_{1g}$ mode
into the acoustic mode, as seen in the top trace in Fig.~\ref{fig:Major-Results} (b). 
In contrast, sudden squeezing oscillations will peak within a quarter cycle after photoexcitation and decay over time.
Second, in a driven parametric oscillator, the drive displacement $A(t)$
is $\pi/4$ out of phase with the amplitude of the signal and idler displacements $u_{\bm{q}}$. 

\begin{figure}
\includegraphics{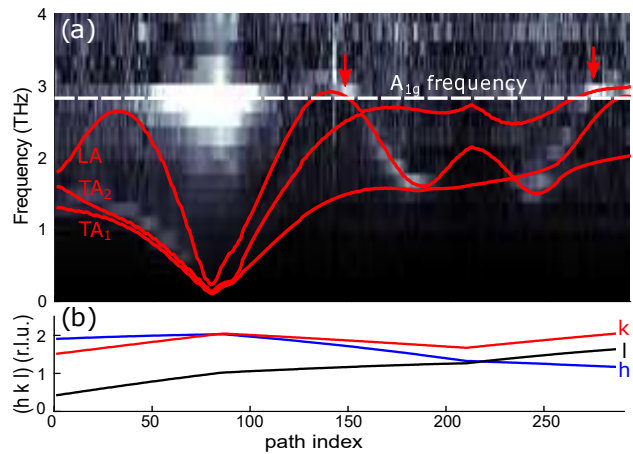}
\caption{\label{fig:Lineouts}(a) Lineout along the cut shown in red in Fig.~\ref{fig:3Thz-image}(a). The intensity is scaled by $\omega^{2}$.
The bright spot near $q=0$ at 2.85 THz is from the $A_{1g}$ mode.
The second harmonic of the acoustic dispersion is plotted on top of
the lineout. The $A_{1g}$ mode frequency is represented by a white
line. The squeezing signal is brightest where the phonon branches
intersect the $A_{1g}$ frequency (red arrows). (b) The reciprocal
lattice positions $(h\ k\ l)$ indices (in r.l.u) along the cut.}
\end{figure}

We measure the mean-square displacements, $\left\langle u_{\bm{q}}^{2}(t)\right\rangle $,
which results in a $\pi/2$ phase shift between $\left\langle u_{\bm{q}}^{2}(t)\right\rangle $ and $A(t)$, as is seen in the phase difference between the traces in Fig.~\ref{fig:Major-Results} (b), in contrast to directly squeezed phonons, which will be in phase with the coherent $A_{1g}$ mode at moderate squeezing amplitude.  Finally, the spectral content of the sudden and resonant squeezed modes are distinct. The parametrically-squeezed modes have the largest oscillations in the mean-square displacements near the resonance $2\omega_{\bm{q}}=\Omega$. 
In contrast, sudden squeezing produces a broad continuum of modes with an amplitude  which is proportional to $k_B T/\omega_{\bm{q}}^{2}$ in the high temperature limit.

In Fig.~\ref{fig:Lineouts} we show the Fourier magnitude along the lineout in reciprocal space given by the red line in Fig.~\ref{fig:3Thz-image} (a).  The Fourier transform amplitude is scaled by $\omega^{2}$ to compensate for the $\propto 1/\omega_{\bm{q}}^{2}$ amplitude of the suddenly squeezed modes. 
The solid red lines in Fig.~\ref{fig:Lineouts}(a) are the second harmonic of the acoustic dispersion relation calculated from DFPT \citep{Murray2007a} and the dashed white line indicates the frequency of the  $A_{1g}$ mode. The calculated frequencies were scaled by approximately 15\% to more closely match the measured position of the resonance. 
Sudden squeezing is responsible for the intensity of the lower frequencies in Fig.~\ref{fig:Lineouts}(a), mostly the TA branches. In contrast, the bright features at 2.85~THz marked with red arrows are due to the parametric resonance.  The corresponding momentum transfer, $\bm{Q}=(h\ k\ l)$, r.l.u. for every point along the lineout is shown in Fig.~\ref{fig:Lineouts}(b), and the resonances occur near $\bm{Q} = (1.71\ 1.86\ 1.16)$ and $\bm{Q} = (1.19\ 2.00\ 1.60)$ r.l.u.

To estimate an acoustic phonon decay rate $\gamma_{\bm{q}}$ for extracting coupling constants, the regions of this lineout with observed frequencies of 1-2 THz were fit to decaying exponentials in the time domain.  The decay rate of the mean-squared fluctuations is $ 0.3\ \mathrm{ps^{-1}}$, which corresponds to an energy decay rate of  $\gamma_{\bm{q}} = 0.3 \ \mathrm{ps^{-1}}$. This decay rate is independent of frequency, and was used as the approximate phonon decay rate $\gamma_{\bm{q}}$ for the LA phonons.

In conclusion, we report the first observation of resonant squeezing
of acoustic phonons by anharmonic coupling to the $A_{1g}$ mode in
bismuth and measure the decay products over a wide range of the Brillouin zone.
We have also measured the anharmonic coupling constants for a representative decay channel within an order of magnitude of first-principles calculations. This work demonstrates a method that can measure individual
phonon-phonon coupling channels throughout the Brillouin zone. These results should be independent of how the coherent motion at zone center is initiated. Of particular interest is the anharmonicity of zone-center IR-active modes that, when driven strongly by mid-IR laser pulses, have been used to manipulate the macroscopic phase of the material \citep{forst2011nonlinear}. This approach to measuring phonons away from zone-center can image the short-range lattice fluctuations that couple to macroscopic changes in electronic properties \citep{Mitrano2016}. Furthermore, this generalizes to situations where the driven zone-center boson is not a phonon but some other excitation that couples to collective excitations deep in the Brillouin zone \citep{Liu2013}.

\begin{acknowledgments}
This work was supported by the U.S. Department of Energy, Office of
Science, Office of Basic Energy Sciences through the Division of Materials
Sciences and Engineering under Contract No. DE-AC02-76SF00515. C.
Uher and T. Bailey acknowledge support from the Department of Energy,
Office of Basic Energy Science under Award \# DE-SC-0008574. Work
at the Tyndall National Institute was supported by Science Foundation
Ireland award 12/IA/1601 and the Irish Research Council GOIPG/2015/2784. Measurements were carried out at the Linac
Coherent Light Source, a national user facility operated by Stanford
University on behalf of the U.S. Department of Energy, Office of Basic
Energy Sciences. Preliminary measurements were performed at the Stanford
Synchrotron Radiation Lightsource (Beamline 7\textendash 2), SLAC
National Accelerator Laboratory. We thank Soo Heyong Lee and Wonhyuk Jo for experimental assistance.
\end{acknowledgments}

\bibliographystyle{unsrt}
\bibliography{Citations_for_Bi_Downconversion.bib}

\begin{thebibliography}{10}

\bibitem{Zhong1995}
W.~Zhong, David Vanderbilt, and K.~M. Rabe.
\newblock {First-principles theory of ferroelectric phase transitions for
  perovskites: The case of BaTiO3}.
\newblock {\em Physical Review B}, 52(9):6301--6312, 1995.

\bibitem{Kittel:ISSP}
Charles Kittel.
\newblock {\em {Introduction to Solid State Physics}}.
\newblock John Wiley \& Sons, Inc., New York, 6th edition, 1986.

\bibitem{Zebarjadi2012}
M.~Zebarjadi, K.~Esfarjani, M.~S. Dresselhaus, Z.~F. Ren, and G.~Chen.
\newblock {Perspectives on thermoelectrics: from fundamentals to device
  applications}.
\newblock {\em Energy Environ. Sci.}, 5(1):5147--5162, 2012.

\bibitem{Debernardi1995}
Alberto Debernardi, Stefano Baroni, and Elisa Molinari.
\newblock Anharmonic phonon lifetimes in semiconductors from density-functional
  perturbation theory.
\newblock {\em Phys. Rev. Lett.}, 75:1819--1822, Aug 1995.

\bibitem{Broido2007}
D.~A. Broido, M.~Malorny, G.~Birner, Natalio Mingo, and D.~A. Stewart.
\newblock Intrinsic lattice thermal conductivity of semiconductors from first
  principles.
\newblock {\em Applied Physics Letters}, 91(23):231922, 2007.

\bibitem{Togo2015}
Atsushi Togo, Laurent Chaput, and Isao Tanaka.
\newblock Distributions of phonon lifetimes in brillouin zones.
\newblock {\em Phys. Rev. B}, 91:094306, Mar 2015.

\bibitem{Brockhouse1955}
B.~N. Brockhouse and A.~T. Stewart.
\newblock Scattering of neutrons by phonons in an aluminum single crystal.
\newblock {\em Phys. Rev.}, 100:756--757, Oct 1955.

\bibitem{lovesey1971theory}
Stephen~William Lovesey.
\newblock {\em Theory of thermal neutron scattering: the use of neutrons for
  the investigation of condensed matter}.
\newblock Clarendon Press, 1971.

\bibitem{squires1978introduction}
Gordon~Leslie Squires.
\newblock {\em Introduction to the Theory of Thermal Neutron Scattering}.
\newblock Courier Corporation, 1978.

\bibitem{Krisch2007}
Michael Krisch and Francesco Sette.
\newblock {\em Inelastic X-Ray Scattering from Phonons}, pages 317--370.
\newblock Springer Berlin Heidelberg, Berlin, Heidelberg, 2007.

\bibitem{burkel2000phonon}
Eberhard Burkel.
\newblock Phonon spectroscopy by inelastic x-ray scattering.
\newblock {\em Reports on Progress in Physics}, 63(2):171, 2000.

\bibitem{Baron2014}
Alfred Q.~R. Baron.
\newblock {\em High-Resolution Inelastic X-Ray Scattering II: Scattering
  Theory, Harmonic Phonons, and Calculations}, pages 1--32.
\newblock Springer International Publishing, Cham, 2014.

\bibitem{Baron2014a}
Alfred Q.~R. Baron.
\newblock {\em High-Resolution Inelastic X-Ray Scattering I: Context,
  Spectrometers, Samples, and Superconductors}, pages 1--68.
\newblock Springer International Publishing, Cham, 2014.

\bibitem{Trigo2013}
M.~Trigo, M.~Fuchs, J.~Chen, M.~P. Jiang, M.~Cammarata, S.~Fahy, D.~M. Fritz,
  K.~Gaffney, S.~Ghimire, A.~Higginbotham, S.~L. Johnson, M.~E. Kozina,
  J.~Larsson, H.~Lemke, A.~M. Lindenberg, G.~Ndabashimiye, F.~Quirin,
  K.~Sokolowski-Tinten, C.~Uher, G.~Wang, J.~S. Wark, D.~Zhu, and D.~A. Reis.
\newblock Fourier-transform inelastic x-ray scattering from time-and
  momentum-dependent phonon-phonon correlations.
\newblock {\em Nature Physics}, 9(12):790--794, oct 2013.

\bibitem{Fahy2016}
Stephen Fahy, \'Eamonn~D. Murray, and David~A. Reis.
\newblock Resonant squeezing and the anharmonic decay of coherent phonons.
\newblock {\em Phys. Rev. B}, 93:134308, Apr 2016.

\bibitem{Cheng1990}
T.~K. Cheng, S.~D. Brorson, a.~S. Kazeroonian, J.~S. Moodera, G.~Dresselhaus,
  M.~S. Dresselhaus, and E.~P. Ippen.
\newblock {Impulsive excitation of coherent phonons observed in reflection in
  bismuth and antimony}.
\newblock {\em Applied Physics Letters}, 57(10):1004, 1990.

\bibitem{Sokolowski-Tinten2003}
Klaus Sokolowski-Tinten, Christian Blome, Juris Blums, Andrea Cavalleri,
  Clemens Dietrich, Alexander Tarasevitch, Ingo Uschmann, Eckhard
  F{\"{o}}rster, Martin Kammler, Michael Horn-von Hoegen, and Dietrich von~der
  Linde.
\newblock {Femtosecond X-ray measurement of coherent lattice vibrations near
  the Lindemann stability limit.}
\newblock {\em Nature}, 422(6929):287--289, mar 2003.

\bibitem{Fritz2007}
D~M Fritz, D~A Reis, B~Adams, and R~A Akre.
\newblock {Ultrafast bond softening in bismuth: Mapping a solid's interatomic
  potential with X-rays}.
\newblock {\em Science}, 315(February):633--637, 2007.

\bibitem{Johnson2009}
S.~L. Johnson, P.~Beaud, E.~Vorobeva, C.~J. Milne, {\'{E}}.~D. Murray, S.~Fahy,
  and G.~Ingold.
\newblock {Directly Observing Squeezed Phonon States with Femtosecond X-Ray
  Diffraction}.
\newblock {\em Physical Review Letters}, 102(17):175503, apr 2009.

\bibitem{Xin1989}
Xin Ma and William Rhodes.
\newblock Squeezing in harmonic oscillators with time-dependent frequencies.
\newblock {\em Phys. Rev. A}, 39:1941--1947, Feb 1989.

\bibitem{Chollet2015}
Matthieu Chollet, Roberto Alonso-Mori, Marco Cammarata, Daniel Damiani, Jim
  Defever, James~T. Delor, Yiping Feng, James~M. Glownia, J.~Brian Langton,
  Silke Nelson, Kelley Ramsey, Aymeric Robert, Marcin Sikorski, Sanghoon Song,
  Daniel Stefanescu, Venkat Srinivasan, Diling Zhu, Henrik~T. Lemke, and
  David~M. Fritz.
\newblock {The X-ray Pump-Probe instrument at the Linac Coherent Light Source}.
\newblock {\em Journal of Synchrotron Radiation}, 22(November 2014):503--507,
  2015.

\bibitem{Herrmann2013}
Sven Herrmann, S{\'{e}}bastien Boutet, Brian Duda, David Fritz, Gunther Haller,
  Philip Hart, Ryan Herbst, Christopher Kenney, Henrik Lemke, Marc
  Messerschmidt, Jack Pines, Aymeric Robert, Marcin Sikorski, and Garth
  Williams.
\newblock {CSPAD-140k : A versatile detector for LCLS experiments}.
\newblock {\em Nuclear Instruments and Methods in Physics Research},
  718:550--553, 2013.

\bibitem{Li2013}
J.~J. Li, J.~Chen, D.~A. Reis, S.~Fahy, and R.~Merlin.
\newblock Optical probing of ultrafast electronic decay in bi and sb with slow
  phonons.
\newblock {\em Phys. Rev. Lett.}, 110:047401, Jan 2013.

\bibitem{Henighan2016}
T.~Henighan, M.~Trigo, M.~Chollet, J.~N. Clark, S.~Fahy, J.~M. Glownia, M.~P.
  Jiang, M.~Kozina, H.~Liu, S.~Song, D.~Zhu, and D.~a. Reis.
\newblock {Control of two-phonon correlations and the mechanism of
  high-wavevector phonon generation by ultrafast light pulses}.
\newblock {\em Physical Review B}, 94(2):020302, jul 2016.

\bibitem{Zhu2015}
Diling Zhu, Aymeric Robert, Tom Henighan, Henrik~T. Lemke, Matthieu Chollet,
  J.~Mike Glownia, David~a. Reis, and Mariano Trigo.
\newblock {Phonon spectroscopy with sub-meV resolution by femtosecond x-ray
  diffuse scattering}.
\newblock {\em Physical Review B}, 92(5):054303, aug 2015.

\bibitem{Murray2007a}
{\'{E}}~Murray, S~Fahy, D~Prendergast, T~Ogitsu, D~Fritz, and D~Reis.
\newblock {Phonon dispersion relations and softening in photoexcited bismuth
  from first principles}.
\newblock {\em Physical Review B}, 75(18):184301, may 2007.

\bibitem{forst2011nonlinear}
M~F{\"o}rst, C~Manzoni, S~Kaiser, Y~Tomioka, Y~Tokura, R~Merlin, and
  A~Cavalleri.
\newblock Nonlinear phononics as an ultrafast route to lattice control.
\newblock {\em Nature Physics}, 7:854--856, 2011.

\bibitem{Mitrano2016}
M.~Mitrano, A.~Cantaluppi, D.~Nicoletti, S.~Kaiser, A.~Perucchi, S.~Lupi,
  P.~Di~Pietro, D.~Pontiroli, M.~Ricc{\`o}, S.~R. Clark, D.~Jaksch, and
  A.~Cavalleri.
\newblock Possible light-induced superconductivity in k3c60 at high
  temperature.
\newblock {\em Nature}, 530(7591):461--464, Feb 2016.
\newblock Letter.

\bibitem{Liu2013}
H.~Y. Liu, I.~Gierz, J.~C. Petersen, S.~Kaiser, A.~Simoncig, A.~L. Cavalieri,
  C.~Cacho, I.~C.~E. Turcu, E.~Springate, F.~Frassetto, L.~Poletto, S.~S.
  Dhesi, Z.-A. Xu, T.~Cuk, R.~Merlin, and A.~Cavalleri.
\newblock Possible observation of parametrically amplified coherent phasons in
  k${}_{0.3}$moo${}_{3}$ using time-resolved extreme-ultraviolet angle-resolved
  photoemission spectroscopy.
\newblock {\em Phys. Rev. B}, 88:045104, Jul 2013.

\end{thebibliography}

\end{document}